# MESOSCOPIC THEORY OF DEFECT ORDERING-DISORDERING TRANSITIONS IN THIN OXIDE FILMS

Anna N. Morozovska[1], Eugene A. Eliseev[2], Dmitry V. Karpinsky[3], Maxim V. Silibin[4a,b], Rama Vasudevan[5], Sergei V. Kalinin[5 *], and Yuri A. Genenko[6†]

*[1]Institute of Physics, National Academy of Sciences of Ukraine,*
*46, pr. Nauky, 03028 Kyiv, Ukraine*

*[2] Institute for Problems of Materials Science, National Academy of Sciences of Ukraine,*
*Krjijanovskogo 3, 03142 Kyiv, Ukraine*

*[3] Scientific-Practical Materials Research Centre of NAS of Belarus, 220072 Minsk, Belarus*

*[4a]National Research University of Electronic Technology "MIET", 124498 Moscow, Russia*

*[4b] Institute for Bionic Technologies and Engineering, I.M. Sechenov First Moscow State Medical University, Moscow 119991, Russia*

*[5]The Center for Nanophase Materials Sciences, Oak Ridge National Laboratory,*
*Oak Ridge, TN 37922*

*[6]Institute of Materials Science, Technische Universität Darmstadt, Otto-Berndt-Str.3,*
*Darmstadt, Germany*

## Abstract

Ordering of mobile defects in functional materials can give rise to fundamentally new phases possessing ferroic and multiferroic functionalities. Here we develop the Landau theory for strain induced ordering of defects (e.g. oxygen vacancies) in thin oxide films, considering both the ordering and wavelength of possible instabilities. Using derived analytical expressions for the energies of various defect-ordered states, we calculated and analyzed phase diagrams dependence on the film-substrate mismatch strain, concentration of defects, and Vegard coefficients. Obtained results open possibilities to create and control superstructures of ordered defects in thin oxide films by selecting the appropriate substrate and defect concentration.

[*] Corresponding author: sergei2@ornl.gov
[†] Corresponding author: genenko@mm.tu-darmstadt.de

## I. INTRODUCTION

Ferroelectric and multiferroic materials are the object of much fascination in physics community, both due to the multitude of possible applications and broad spectrum of fundamental physical phenomena they exhibit. Applications such as ferroelectric memories, field effect transistors, and domain wall conductance have riveted attention of scientific community in the last two decades [1, 2, 3]. Similarly, the nature of ferroelectric transitions, ferroelectricity in disordered systems, etc. remain a subject of active research since the discovery of ferroelectricity in late 1920ies [4]. Topological defects in ferroelectric materials and coupling between the ferroelectric and semiconductor subsystems are actively explored in the context of surface and domain wall conductance [5, 6, 7, 8].

From the gamut of possible behaviors, in the last decade progressively more attention is focused on multiferroic materials, i.e. systems possessing two or more order parameters [9, 10, 11, 12]. These functionalities significantly broaden the spectrum of possible applications, including oxide nanoelectronics, sensors and actuators, and IoT devices [13, 14, 15, 16]. Furthermore, in multiferroic materials fundamentally new properties can emerge at the topological defects [17]. Examples of the former include the emergence of suppressed order parameter and associated topological defects at the domain walls, surfaces and interfaces, domain wall and vortex core conduction, incipient ferroelectricity and magnetism, and many others [8, 18, 19]. Remarkedly that conductive domain walls in bulk samples and thin films can act as mobile charged channels (being also mobile topological defects) opening the way for "domain wall nanoelectronics" [20]. Additional degrees of functionality of nanosized multiferroics with the topological defects [21, 22] and newly discovered low-dimensional semiconductor materials come from the strain engineering and straintronics concept [23].

The vast majority of the research in the field explored the coupling between the primary physical order parameters including magnetization, polarization, and ferroelasticity [8, 18]. However, it is well known that chemical degrees of freedom can strongly affect the ferroic behavior [24]. A number of groups explored the phenomena such as vacancy segregation at the domain walls and grain boundaries [25, 26, 27, 28, 29], or changes in surface reactivity in response to polarization [30, 31]. However, most of these works explore the responses of chemical subsystems to the polarization. Relatively small effort was dedicated to the exploration of chemical effects on polarization [32, 33, 34], typically the surface electrochemistry including chemically-induced switching [35, 36, 37, 38] and emergence of ferroionic phases [39, 40, 41]. This direction has acquired particular prominence with the advent of the hybrid perovskites where chemical subsystem is strongly coupled to the environment and electrode phenomena (see e.g. review [42] and refs therein). However, the volume of research in this field is limited.

Here, we develop an initial framework for mesoscopic theory of defect ordering-disordering transitions in thin oxide films. We explore whether chemically induced changes in Vegard volume can be used to trigger and control multiferroic orderings in thin strained films (see e.g. Ref. [21-23] and refs. therein), and what are the properties of such systems.

## II. TWO SUB-LATTICES MODEL OF POINT DEFECTS ORDERING

Here we consider the case of a thin oxide film on a substrate, taking into account the misfit strain [43], while neglecting the appearance of misfit dislocations. This assumption is generally valid for the films with a thickness smaller than a critical value (~ 10 - 50 nm) corresponding to dislocation nucleation [44]. The geometry of the problem is shown in **Fig. 1.** For high enough concentration of neutral/charged vacancies, the vacancy-ordered state can emerge [45] leading to the appearance of an elastic/electric dipole sublattice.

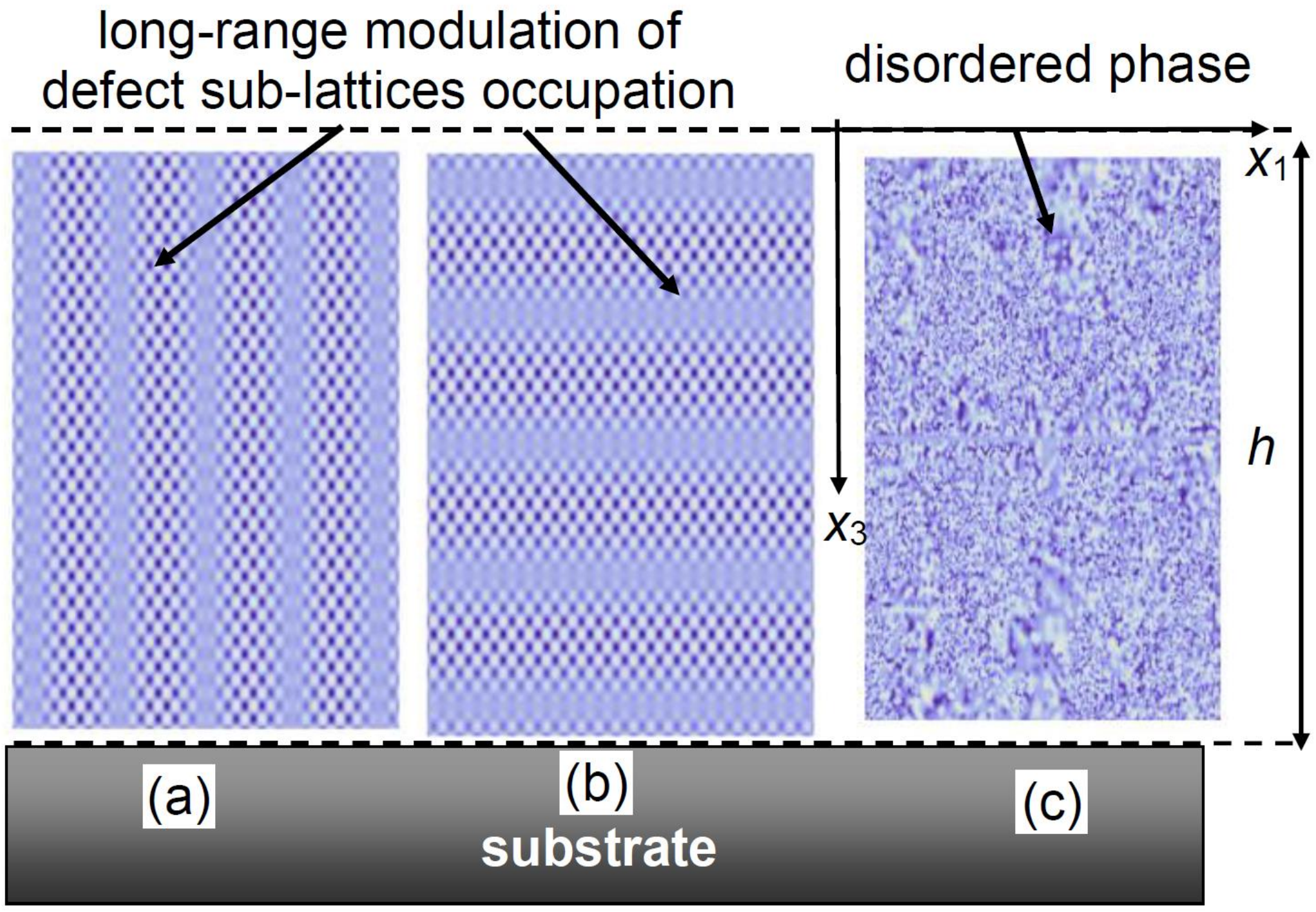

**FIGURE 1.** Point defect ordering in ultra-thin oxide film of thickness *h* placed on a rigid substrate. One can distinguish perpendicular (**a**, the left side) and parallel (**b**, the central part) orientations of the defect ordering

from the state with randomly distributed defects (**c**, the right side). Parts (**a)** and (**b)** illustrate how the harmonic modulations of the order parameter reveal local redistribution of defects between defect sublattices.

Rigorously speaking, in complex oxides there may be more than two oxygen sublattices [46]. However, our simplified continuum model with a scalar order parameter is naturally limited to two sublattices. The suggested model is isotropic in the $\{x_1, x_2\}$ plane together with the misfit strain. Since the symmetry breaking may spontaneously occur along any in-plane direction, this model is able to describe, particularly, a possible oxygen vacancy ordering along the $x_1$ or the $x_2$ axis. Thanks to the model symmetry, it is sufficient to consider exemplary cases of ordering along the $x_1$ axis, corresponding to the in-plane order parameter modulation, and along the $x_3$ axis, corresponding to the order parameter modulation perpendicular to the film surfaces.

The spatial scale of the defect concentration fluctuations, which can be correctly considered in a continuum approach, should be much larger than the lattice constant (so called long-range fluctuations). Thus, to describe the phase ordering process, we introduce a dimensionless long-range order parameter $\eta$, related to the degree of ordering of point defects (oxygen or cation vacancies, or impurity atoms) and given by a disbalance of occupations $\delta n_a$ and $\delta n_b$ of the two sub-lattices: $\eta = \frac{c}{c_0}(\delta n_b - \delta n_a)$. Here $c$ is the concentration of defects related with the non-stoichiometry degree δ; $c_0$ is a characteristic value that usually coincides with the solubility limit, meaning that the defect atoms do not affect essentially the host lattice force matrix at $c < c_0$. As one can see, the scalar order parameter η is a normalized occupation degree of the sub-lattices, $0 \leq |\eta| \leq 1$. For given η, the relative occupation numbers for the two sub-lattices are $\delta n_a = c(1-\eta)/(2c_0)$ and $\delta n_b = c(1+\eta)/(2c_0)$. The total occupation of two sub-lattices is constant, $\delta n_a + \delta n_b = c/c_0$. Values $\eta = \pm 1$ correspond to the complete ordering of defects in either sub-lattice "b" or "a", while $\eta = 0$ corresponds to the complete disorder (with equal filling of two sub-lattices).

Note that $c_0 < N_a$, where $N_a$ is the stoichiometric concentration of host atoms. In the particular case of oxygen vacancies, $c$ depends on the oxygen pressure $p$ via the defect equilibria and for certain cases $\frac{c}{c_0} \sim \delta \sim ln\left(\frac{p_0 + A(T)p}{p_0}\right)$ (e.g. see Fig. 3 in Ref. [47]). Following the theory by Khachaturyan [48], we assume that concentration $c$ is coordinate-independent [49].

The defect ordering direction can be different with respect to the film surfaces. Below we will distinguish parallel [see the left side **(a)** in **Fig. 1**] and perpendicular [see the central part **(b)** in **Fig. 1**] orientations of the planes of a constant order parameter and, thus, of equal sublattice occupations.

The η-dependent free energy of a thin film is the sum of the surface and bulk energies:

$$F = F_S + F_V, \quad F_S = \frac{\alpha_S}{2}\int_{S_F}[\eta^2(x_1,x_2,0) + \eta^2(x_1,x_2,h)]dx_1dx_2, \; F_V = \int_{V_F} f dx_1dx_2dx_3. \quad \text{(1a)}$$

where $h$ is the film thickness, and the first term is the surface/interface energy that is not negative under the condition $\alpha_S \geq 0$. Hereinafter we will consider the case $\alpha_S = 0$, corresponding to the so-called natural boundary conditions.

The bulk density, *f*, of the Helmholtz free energy dependence on the order parameter $\eta$ has the form:

$$f(x_1,x_2,x_3) = c^2\frac{\alpha}{2}\eta^2 + \frac{c\,k_BT}{2}[(1-\eta)\ln(1-\eta) + (1+\eta)\ln(1+\eta)] + c^2\frac{g_{kl}}{2}\frac{\partial\eta}{\partial x_k}\frac{\partial\eta}{\partial x_l} +$$
$$c^2\frac{w_{ijkl}}{2}\frac{\partial^2\eta}{\partial x_i\partial x_j}\frac{\partial^2\eta}{\partial x_k\partial x_l} + q\left[\eta,\sigma_{ij}\right] \quad \text{(1b)}$$

The first term in Eq. (1b), proportional to $\eta^2$, originates from the interaction between defects from the different sublattices. The term can be readily derived from the term $\delta n_a\,\delta n_b$ where $\delta n_a$ and $\delta n_b$ are the relative occupation numbers for the two sub-lattices. Hereinafter we will consider both cases $\alpha < 0$ and $\alpha > 0$. For the former, defects may order spontaneously in a bulk material, while in the latter case the ordering in bulk is unfavorable, but can be induced by external factors (such as misfit strain).

The second term in Eq. (1b) is proportional to the entropy of the system with a minus sign, $-TS$, [42, 44], since $F = U - TS$. The relative dimensionless entropy, defined as $-(1-\eta)\ln(1-\eta) - (1+\eta)\ln(1+\eta)$, tends to the minimum of $-2\ln 2$ under the condition $\eta^2 \to 1$, corresponding to the complete occupation of one of the sublattices ($\delta n_b = 1$ and $\delta n_a = 0$, or vice versa $\delta n_a = 1$ and $\delta n_b = 0$).We note that the entropy contribution stabilizes the system thermodynamically since its expansion contains all positive even powers of the order parameter.

The third and the fourth terms are the series expansion of generalized gradient energy in the derivatives $\partial\eta/\partial x_i$ of the order parameter. The gradient energy was introduced by Cahn and Hillard in 1958 (see e.g. [50]), since its inclusion becomes indispensable for correct description of inhomogeneous electrochemical systems with defect species. Below we consider the tensors $g_{ij}$ and $w_{ijkl}$ in isotropic approximation, $g_{ij} = g\delta_{ij}$ and $w_{ijkl} = w\delta_{ij}\delta_{kl}$, and explore the case of $w > 0$. This choice is because we study the case of the defects ordering for positive correlation energy.

Note that very often the higher gradient term is neglected ($w = 0$). Rigorously it is justified only if $g > 0$, assuming that its renormalization by the strain gradient energy is either positive or too small to change the positive sign of resulting effective gradient coefficient. Below we will show that $w$-term is mandatory to determine the threshold for emergence of the order parameter spatial modulation.

The elastic energy $q\left[\eta, \sigma_{ij}\right]$ includes the coupling between the order parameter and the stresses $\sigma_{ij}$. Following the theoretical formalism of Vegard strains, proposed by Freedman [51], and using the strain energy by Levanyuk et al. [52], the coupling term between the order parameter and the stresses could be also expanded as a series on the order parameter:

$$q\left[\eta, \sigma_{ij}\right] = -\frac{s_{ijkl}}{2}\sigma_{ij}\sigma_{kl} + u_{ij}\sigma_{ij} - c_0\sigma_{ij}\left(V_{ij}^{(a)}\delta n_a + V_{ij}^{(b)}\delta n_b\right) - c^2\frac{B_{ijkl}}{2}\sigma_{ij}\frac{\partial\eta}{\partial x_k}\frac{\partial\eta}{\partial x_l}$$

$$= -\frac{s_{ijkl}}{2}\sigma_{ij}\sigma_{kl} + u_{ij}\sigma_{ij} - c\left(V_{ij}^{c} + \eta V_{ij}^{\eta}\right)\sigma_{ij} - c^2\frac{B_{ijkl}}{2}\sigma_{ij}\frac{\partial\eta}{\partial x_k}\frac{\partial\eta}{\partial x_l}. \quad (1c)$$

Here $s_{ijkl}$ is the elastic compliances tensor, and $u_{ij}$ is the tensor of elastic strains. To describe the mixed mechanical state of the epitaxial binary oxide film on the substrate, we perform the Legendre transform by adding the term $u_{kl}\sigma_{kl}$. The third term in Eq. (1c) is the chemical expansion due to the appearance of elastic defects, $c_0\sigma_{ij}\left(V_{ij}^{(a)}\delta n_a + V_{ij}^{(b)}\delta n_b\right)$, where the tensors $V_{ij}^{(a)}$ and $V_{ij}^{(b)}$ characterize the Vegard strains for defects located within the sublattices (a) and (b), respectively. The symmetry of $V_{ij}^{(a,b)}$ is determined by the local site symmetry of defect with respect to the lattice, which could be different from the lattice symmetry (see e.g. Ref. [51] considering different vacancies in $SrTiO_3$). Using the expressions of the partial sublattice occupation numbers through the order parameter $\eta$ the third term in Eq. (1c) can be further transformed to $c\left(V_{ij}^{c} + \eta V_{ij}^{\eta}\right)$ with the mean value $V_{ij}^{c} = \left(V_{ij}^{(a)} + V_{ij}^{(b)}\right)/2$ and the difference $V_{ij}^{\eta} = \left(V_{ij}^{(b)} - V_{ij}^{(a)}\right)/2$.

It will be shown that the last term in Eq. (1c) plays a central role in the mechanism of defect ordering. It is a gradient-type striction due to the defect ordering that is characterized by a fourth rank tensor $B_{ijkl}$.

In the continuous media approximation, the thermodynamically stable state of the film can be derived from the variation of the free energy (1) on η and $\sigma_{ij}$, leading to the Euler-Lagrange differential equations:

$$\alpha c^2\eta + ck_BT\,\text{arctanh}(\eta) + c^2B_{ijkl}\frac{\partial\sigma_{ij}}{\partial x_k}\frac{\partial\eta}{\partial x_l} - c^2\left(g_{kl} - B_{ijkl}\sigma_{ij}\right)\frac{\partial^2\eta}{\partial x_k\partial x_l} + \frac{c^2 w_{ijkl}\partial^4\eta}{\partial x_i\partial x_j\partial x_k\partial x_l} - c\,V_{ij}^{\eta}\sigma_{ij} = 0 \qquad (2a)$$

$$c\left(V_{ij}^{c} + \eta\,V_{ij}^{\eta}\right) + c^2\frac{B_{ijkl}}{2}\frac{\partial\eta}{\partial x_k}\frac{\partial\eta}{\partial x_l} + s_{ijkl}\sigma_{kl} = u_{ij} \qquad (2b)$$

Here we used the identity, $\text{arctanh}(\eta) \equiv \frac{1}{2}ln\left(\frac{1+\eta}{1-\eta}\right)$.

The boundary conditions to Eqs. (2a) are the following:

$$\left[\alpha_S\eta \pm gc^2\frac{\partial\eta}{\partial z} \mp B_{ij33}\sigma_{ij}c^2\frac{\partial\eta}{\partial z}\right]\Big|_{x_3=0,h} = 0. \qquad (2c)$$

The first term in Eq.(2c) originated from the variation of surface energy in Eq. (1a), and the so-called natural boundary conditions correspond to the case $\alpha_S = 0$ (surface energy is absent in this particular case). The second term in Eq. (2c) originated from the variation of the gradient energy in Eq. (1b), and the third term originated from the variation of the elastic energy Eq. (1c).

Eq. (2b) should be considered along with the conditions of mechanical equilibrium $\partial\sigma_{ij}/\partial x_j = 0$. The elastic boundary conditions to Eqs. (2b) are the following:

$$\sigma_{i3}(x_1, x_2, 0) = 0, \quad u_{11}(x_1, x_2, h) = u_{22}(x_1, x_2, h) = u_m. \qquad (2d)$$

The first condition, $\sigma_{i3} = 0$, means that the top surface of the film ($x_3 = 0$) is mechanically free; and the second condition, $u_{11} = u_{22} = u_m$, means that the bottom surface of the film ($x_3 = h$) is clamped to a rigid substrate, where $u_m$ is a misfit strain induced by the film-substrate lattices mismatch.

Below we consider thicknesses $h$ smaller than the critical thickness of misfit dislocation appearance [44], typically this means that $h \leq$ 10 nm.

The nonlinear boundary problem (2) contains a number of material-dependent constants, the majority of which are poorly known even for simple binary oxides, such as ZnO, MgO, $SnO_2$, $CeO_2$, $HfO_2$, and even more so for complex ternary oxides, such as manganites (La,Sr)$MnO_3$, paraelectric perovskites $SrTiO_3$, $EuTiO_3$, $KTaO_3$, ferroelectric perovskites (Ba,Sr)$TiO_3$, (Pb,Zr)$TiO_3$, and orthoferrites $PbFeO_3$, pristine and rare-earth doped $BiFeO_3$, etc., which all can be deficient in oxygen. Hence, prior to solving the boundary problem (2) by e.g. finite element modeling (**FEM**), requiring all tabulated parameters, these should be taken from the experimental or density functional studies [5, 6, 8, 19]. However, to facilitate the search in the multi-parameter space and open the way for further FEM, here we elaborate the analytical theory.

## III. ANALYTICAL SOLUTION IN HARMONIC APPROXIMATION

To explore the phase evolution in the system described by the free energy, Eq. (1), we consider the three-dimensional Fourier series of the order parameter,

$$\eta(\boldsymbol{x}) = \sum_{l,m,n=-\infty}^{\infty} \tilde{\eta}_{lmn}\, exp\left[i\left(k_1^{(lmn)}x_1 + k_2^{(lmn)}x_2 + k_3^{(lmn)}x_3\right)\right], \tag{3a}$$

where the wave vector components $\boldsymbol{k}^{lmn} = \left(\frac{2\pi}{L}l, \frac{2\pi}{L}m, \frac{2\pi}{h}n,\right)$ are determined by the sizes $L \times L \times h$ of the considered film, and *m*, *n*, and *l* are integer numbers. For the order parameter $\eta(\boldsymbol{x})$ to be a real (i.e. observable) value, the equality $\tilde{\eta}^*_{l,m,n} = \tilde{\eta}_{-l,-m,-n}$ should be valid.

Following Landau theory, below we assume that only one principal mode **k** dominates near the order-disorder phase transition,

$$\eta(\boldsymbol{x}) \cong \eta_0 + \tilde{\eta}\exp(i\boldsymbol{kx}) + c.c. \equiv \eta_0 + 2|\tilde{\eta}|\cos[(\boldsymbol{kx} + \delta)], \tag{3b}$$

where δ is the phase shift and $\eta_0$ is the offset term. Next, we assume that the approximate equality is valid for the order parameter derivative in Eq. (1c), namely:

$$\frac{\partial\eta}{\partial x_m}\frac{\partial\eta}{\partial x_n} \cong 2k_m k_n|\tilde{\eta}|^2(1 - \cos[2(\boldsymbol{kx} + \delta)]) \tag{3c}$$

Note that **k** is constant in the first harmonic approximation, but should be found in the self-consistent way in what follows.

Substitution of Eqs. (3) in Eqs. (1b) and (1c) yields:

$$\frac{g_{ij}c^2}{2}\frac{\partial\eta}{\partial x_i}\frac{\partial\eta}{\partial x_j} \cong c^2\tilde{g}|\tilde{\eta}|^2(1 - \cos[2(\boldsymbol{kx} + \delta)]), \tag{4a}$$

$$\frac{B_{ijmn}}{2}\sigma_{ij}c^2\frac{\partial\eta}{\partial x_m}\frac{\partial\eta}{\partial x_n} \cong c^2\tilde{B}_{ij}(\boldsymbol{k})\sigma_{ij}|\tilde{\eta}|^2(1 - \cos[2(\boldsymbol{kx} + \delta)]), \tag{4b}$$

$$\frac{w_{ijkl}c^2}{2}\frac{\partial^2\eta}{\partial x_i\partial x_j}\frac{\partial^2\eta}{\partial x_k\partial x_l} \cong c^2\tilde{w}|\tilde{\eta}|^2(1 + \cos[2(\boldsymbol{kx} + \delta)]), \tag{4c}$$

where $\tilde{g}(\boldsymbol{k}) = g_{ij}k_ik_j$, $\tilde{B}_{ij}(\boldsymbol{k}) = B_{ijmn}k_mk_n$ and $\tilde{w}(\boldsymbol{k}) = w_{ijmn}k_ik_jk_mk_n$.
Symmetry of the gradient-striction tensor $B_{ijkl}$ is determined by the local site symmetry of defect with respect to the lattice. Below we assume the tetragonal symmetry of the tensor $\tilde{B}_{ij}(\boldsymbol{k})$, with a tetragonal axis along $x_3$ axis of the film.

In the continuum approximation, the thermodynamically stable state of the film can be analyzed by the variation of the free energy (1) that acquires the form:

$$F = \int_{V_F} d\boldsymbol{x}\Big[\frac{\alpha}{2}c^2\eta^2 + c\frac{k_BT}{2}[(1-\eta)\ln(1-\eta) + (1+\eta)\ln(1+\eta)] - c\left(V_{ij}^c + V_{ij}^\eta(\eta_0 + 2|\tilde{\eta}|\cos[(\boldsymbol{kx}+\delta)])\right)\sigma_{ij} - \frac{s_{ijkl}}{2}\sigma_{ij}\sigma_{kl} + u_{kl}\sigma_{kl} + c^2\tilde{g}(\boldsymbol{k})|\tilde{\eta}|^2(1 - \cos[2(\boldsymbol{kx}+\delta)]) + c^2\tilde{w}(\boldsymbol{k})|\tilde{\eta}|^2(1 + \cos[2(\boldsymbol{kx}+\delta)]) - c^2\tilde{B}_{ij}(\boldsymbol{k})\sigma_{ij}|\tilde{\eta}|^2(1 - \cos[2(\boldsymbol{kx}+\delta)])\Big] \tag{5a}$$

As a next step, one can use a Galerkin procedure in the free energy (5a) with respect to the second harmonics assuming statistical averaging, equivalent to the spatial averaging in ergodic case. Such averaging "excludes" the contributions from the second and higher harmonics to $\cos[2(\boldsymbol{kx}+\delta)]^{2n+1}$ due to a periodicity of trigonometric functions. Assuming the absence of correlations between the offset term $\eta_0$ and $\cos[2(\boldsymbol{kx}+\delta)]$, all cross-terms proportional to $\eta_0^{2m+1}\cos[2(\boldsymbol{kx}+\delta)]^{2n+1}$ vanish after the averaging. Next, the terms proportional to $\eta_0^{2m+1}$ also vanish, while the even terms $\eta_0^{2m}$ do not contain any useful information about defect ordering. Thus, one can regard that $\langle\eta^2\rangle\sim|\tilde{\eta}|^2\sim\eta^2$, where the brackets $\langle\dots\rangle$ designate the averaging, and obtain:

$$\langle f\rangle\sim\frac{\alpha}{2}c^2\eta^2+c\frac{k_BT}{2}[(1-\eta)\ln(1-\eta)+(1+\eta)\ln(1+\eta)]-c\left(V_{ij}^c+\eta_0V_{ij}^{\eta}\right)\sigma_{ij}-\frac{s_{ijkl}}{2}\sigma_{ij}\sigma_{kl}+u_{kl}\sigma_{kl}+c^2\tilde{g}(\boldsymbol{k})|\tilde{\eta}|^2+c^2\tilde{w}(\boldsymbol{k})|\tilde{\eta}|^2-c^2\tilde{B}_{ij}(\boldsymbol{k})\sigma_{ij}|\tilde{\eta}|^2 \tag{5b}$$

Minimization of (5b), $\frac{\partial\langle f\rangle}{\partial\eta}=0$ and $\frac{\partial\langle f\rangle}{\partial\sigma_{ij}}=0$, instead of the Euler-Lagrange Eqs. (2), yields the algebraic equations of state:

$$c\left(\alpha+2\tilde{g}+2\tilde{w}-2\tilde{B}_{ij}\sigma_{ij}\right)\eta+k_BT\,\mathrm{arctanh}(\eta)=0, \tag{6a}$$

$$c\left(V_{ij}^c+\eta_0\,V_{ij}^{\eta}\right)+\tilde{B}_{ij}c^2\eta^2+s_{ijkl}\sigma_{kl}=u_{ij}, \tag{6b}$$

Near the phase transition, $\mathrm{arctanh}(\eta)\approx\eta+\frac{\eta^3}{3}$ in Eq. (6a).

Assuming that the anisotropic Vegard tensor is diagonal (or at least can be diagonalized), that is true for many cases [51], one can write $V_{ij}^c=V_{ii}^c\delta_{ij}$ and $V_{ij}^{\eta}=V_{ii}^{\eta}\delta_{ij}$. Using this approximation, elastic solution for a thin oxide film on a rigid substrate is derived for cubic symmetry far from the structural domain walls. Zero components are $\sigma_{33}=\sigma_{13}=\sigma_{23}=\sigma_{12}=0$, $u_{12}=u_{13}=u_{23}=0$, and nonzero components are:

$$\sigma_{11}=\frac{u_m}{s_{11}+s_{12}}-\frac{s_{11}\left(V_{11}^c+\eta_0V_{11}^{\eta}+\tilde{B}_{11}c\,\eta^2\right)-s_{12}\left(V_{22}^c+\eta_0V_{22}^{\eta}+\tilde{B}_{22}c\,\eta^2\right)}{s_{11}^2-s_{12}^2}c, \tag{7a}$$

$$\sigma_{22}=\frac{u_m}{s_{11}+s_{12}}-\frac{s_{11}\left(V_{22}^c+\eta_0V_{22}^{\eta}+\tilde{B}_{22}c\,\eta^2\right)-s_{12}\left(V_{11}^c+\eta_0V_{11}^{\eta}+\tilde{B}_{11}c\,\eta^2\right)}{s_{11}^2-s_{12}^2}c, \tag{7b}$$

$$u_{11}=u_{22}=u_m, \tag{7c}$$

$$u_{33}=\left(V_{33}^c+\eta_0V_{33}^{\eta}\right)c+\tilde{B}_{33}c^2\eta^2+\frac{s_{12}}{s_{11}+s_{12}}\left[2u_m-\left(V_{11}^c+V_{22}^c+\eta_0V_{11}^{\eta}+\eta_0V_{22}^{\eta}\right)c-\left(\tilde{B}_{11}+\tilde{B}_{22}\right)c^2\eta^2\right]. \tag{7d}$$

The convolutions $\tilde{B}_{11}=B_{11}k_1^2+B_{12}(k_2^2+k_3^2)$, $\tilde{B}_{22}=B_{11}k_2^2+B_{12}(k_1^2+k_3^2)$ and $\tilde{B}_{33}=B_{11}k_3^2+B_{12}(k_1^2+k_2^2)$ are included in Eqs. (7), and the Voight notations for $B_{1111}\equiv B_{11}$ and $B_{1122}\equiv B_{12}$ are used hereinafter.

In a freestanding film the stresses are zero $\sigma_{ij} = 0$, and elastic strains are $u_{ij} = c\left(V_{ij}^{c} + \eta_0 V_{ij}^{\eta}\right) + \tilde{B}_{ij} c^2 \eta^2$.

Below we will suppose that $\eta_0 = 0$ and introduce the designations

$$V_m = V_{11}^{c} + V_{22}^{c}, \qquad V_n = V_{11}^{c} - V_{22}^{c}, \tag{7e}$$

where the sum $V_m$ has the sense of partial molar volume, and the difference $V_n$ reflects the anisotropy impact.

## IV. DEFECT ORDER-DISORDER TRANSITION

### A. Long-range ordering parallel to the film surfaces

Here we consider the case when the only nonzero component of wave vector is $k_3$, meaning that the harmonic modulation of the long-range order parameter $\eta \sim |\tilde{\eta}| \cos[k_3 x_3 + \delta]$ looks like planes parallel to the film surfaces $x_3$=0, $h$ (see **Fig. 3a**, left). Since $\eta$ is proportional to the difference of defect sub-lattice occupations $\delta n_a - \delta n_b$, this means the modulation of the sub-lattice occupation perpendicular to the film surfaces.

For the case the free energy density Eq. (1) has the following form (see Appendix A):

$$f[\eta] = \alpha_p + \frac{\beta_{pr}}{2} c^2 \eta^2 + \frac{\gamma_{pr} c^4}{12} \eta^4 + c \frac{k_B T}{2} [(1-\eta)\ln(1-\eta) + (1+\eta)\ln(1+\eta)] \tag{8a}$$

Where the coefficients $\alpha_p = \frac{(2u_m - V_m c)^2}{4(s_{11}+s_{12})} + \frac{c^2 V_n^2}{4(s_{11}-s_{12})}$, $\beta_{pr} = \alpha + 2(\tilde{g} + \tilde{w}) - 2\tilde{B}_{22} \frac{2u_m - V_m c}{s_{11}+s_{12}}$, $\gamma_{pr} = \frac{12\tilde{B}_{22}^2}{s_{11}+s_{12}}$, and $\tilde{B}_{22} = B_{12} k_3^2$, $\tilde{g} = g k_3^2$ and $\tilde{w} = w k_3^4$. We refer the case with subscript "**pr**".

Omitting the η–independent term $\alpha_{pr}$, the free energy density (8a) can be expanded in power series in η as

$$\delta f[\eta] \approx \left(\beta_{pr} c^2 + c k_B T\right) \frac{\eta^2}{2} + \left(\frac{\gamma_{pr} c^4 + c k_B T}{3}\right) \frac{\eta^4}{4}. \tag{8b}$$

The derivation of Eqs. (8) is given in **Appendix A** [53]. Note that the renormalized coefficient before $\eta^4$ should be positive to ensure the stability of the phase described by the free energy (8b), otherwise higher terms should be included in the expansion. Since $s_{11} > |s_{12}|$ for all elastically stable solids, and $\tilde{B}_{22}^2 \geq 0$, the condition $\gamma_{pr} \geq 0$ is always valid.

The equilibrium values of the order parameter obtained from minimization of the energy (8b) and the corresponding free energy have the form:

$$\eta_S^{pr} = \pm\sqrt{-3 \frac{\beta_{pr} c + k_B T}{\gamma_{pr} c^3 + k_B T}}, \qquad \delta f_{pr}\left[\eta_S^{pr}\right] = -\frac{3c}{4} \frac{\left(\beta_{pr} c + k_B T\right)^2}{\gamma_{pr} c^3 + k_B T}. \tag{8c}$$

Long-range order exists if $\beta_{pr}c + k_B T < 0$.

### B. Long-range ordering perpendicular to the film surface

We further consider the case when the only nonzero component of the wave vector is $k_1$, , meaning that the harmonic modulation of the long-range order parameter $\eta \sim |\tilde{\eta}| \cos[k_1 x_1 + \delta]$ looks like planes perpendicular to the film surfaces $x_3$=0, $h$ (see **Fig. 3a**, middle). Since $\eta$ is proportional to the difference of defect sub-lattice occupations $\delta n_a - \delta n_b$, this means the modulation of the sub-lattice occupations parallel to the film surfaces.

The free energy density $f$ in Eq. (1b) of the oxide has the following form in this case:

$$f[\eta] = \alpha_p + \frac{\beta_{pp}}{2} c^2 \eta^2 + \frac{\gamma_{pp} c^4}{12} \eta^4 + c \frac{k_B T}{2} [(1-\eta)\ln(1-\eta) + (1+\eta)\ln(1+\eta)] \qquad (9a)$$

Where $\beta_{pp} = \alpha + 2\left(\tilde{g} + \tilde{w} - \left(\tilde{B}_{11} + \tilde{B}_{22}\right)\frac{2u_m - V_m c}{2(s_{11}+s_{12})} + \left(\tilde{B}_{11} - \tilde{B}_{22}\right)\frac{V_n c}{2(s_{11}-s_{12})}\right)$ and $\gamma_{pp} = 6\frac{(\tilde{B}_{11}^2 + \tilde{B}_{22}^2)s_{11} - 2\tilde{B}_{11}\tilde{B}_{22}s_{12}}{(s_{11}^2 - s_{12}^2)}$ with $\tilde{B}_{11} = B_{11}k_1^2$, $\tilde{B}_{22} = B_{12}k_1^2$, $\tilde{g} = gk_1^2$ and $\tilde{w} = wk_1^4$. We denote this case with subscript "**pp**".

Omitting the η–independent term $\alpha_p$, the free energy density (9a) can be expanded in powers of η as

$$\delta f[\eta] \approx \left(\beta_{pp}c^2 + ck_B T\right)\frac{\eta^2}{2} + \left(\frac{\gamma_{pp}c^4 + ck_B T}{3}\right)\frac{\eta^4}{4}. \qquad (9b)$$

Note that the renormalized coefficient before $\eta^4$ should be positive, otherwise higher terms should be included in the expansion. Thus, the inequality $\gamma_{pp}c^4 + ck_B T > 0$ should be verified. Since $s_{11} > |s_{12}|$ for all elastically stable solids, and $\tilde{B}_{11}^2 + \tilde{B}_{22}^2 \geq 2\left|\tilde{B}_{11}\tilde{B}_{22}\right|$ the condition $\gamma_{pp} > 0$ is always valid, and so $\gamma_{pp}c^4 + ck_B T > 0$ at finite temperatures.

The equilibrium values of the order parameter obtained from minimization of the energy (9b) and corresponding energy have the form similar to Eq. (8c):

$$\eta_S^{pp} = \pm\sqrt{-3\frac{\beta_{pp}c + k_B T}{\gamma_{pp}c^3 + k_B T}}, \qquad \delta f_{pp}\left[\eta_S^{pp}\right] = -\frac{3c}{4}\frac{\left(\beta_{pp}c + k_B T\right)^2}{\gamma_{pp}c^3 + k_B T}. \qquad (9c)$$

Long-range order exists if $\beta_{pp}c + k_B T < 0$. The derivation of Eqs. (9) is given in **Appendix** [53].

Note that, in the ordered phases, elastic dipoles may become polar due to the surface-induced piezoelectric coupling that, in turn, originates from the inversion symmetry breaking near the film surfaces. At the same time, the out-of-plane polar phase is affected by the strong depolarization field originated from the sharp gradient of polarization decay away from the surfaces.

### C. Structural phase diagrams

The temperatures of transitions between defect-disordered and defect-ordered phases for **parallel** and **perpendicular** orientations of the planes of a constant order parameter can be determined from equations

$$T_{pp}[u_m, c, k_1] = -\frac{\beta_{pp}c}{k_B} \equiv -\frac{c}{k_B}\left(\alpha + 2(gk_1^2 + wk_1^4) - k_1^2\frac{(B_{11}+B_{12})}{s_{11}+s_{12}}(2u_m - V_m c) + k_1^2\frac{V_n(B_{11}-B_{12})}{s_{11}-s_{12}}c\right), \quad (10a)$$

$$T_{pr}[u_m, c, k_3] = -\frac{\beta_{pr}c}{k_B} \equiv -\frac{c}{k_B}\left(\alpha + 2(gk_3^2 + wk_3^4) - \frac{2k_3^2 B_{12}(2u_m - V_m c)}{s_{11}+s_{12}}\right). \quad (10b)$$

We further explore the equilibrium period of $\eta$-modulation, related to the wave vector $\boldsymbol{k}$. In equilibrium the conditions $\frac{\partial T_{pp}}{\partial k_i} = \frac{\partial T_{pr}}{\partial k_i} = 0$ should be fulfilled [54]. Thus $\frac{\partial T_{pp}}{\partial k_1} = -\frac{2c}{k_B}k_1\left(2g - \frac{(B_{11}+B_{12})(2u_m - V_m c)}{s_{11}+s_{12}} + \frac{V_n(B_{11}-B_{12})}{s_{11}-s_{12}}c + 4wk_1^2\right) = 0$ for the constant $\eta$-planes perpendicular to the film surfaces, and $\frac{\partial T_{pr}}{\partial k_3} = -\frac{2c}{k_B}k_3\left(2g - 2B_{12}\frac{2u_m - V_m c}{s_{11}+s_{12}} + 4wk_3^2\right) = 0$ for the constant $\eta$-planes parallel to the film surfaces.

From these conditions the components of the modulation vector acquire the form:

$$k_1 = \pm\sqrt{-\frac{1}{4w}\left(2g - (B_{11}+B_{12})\frac{2u_m - cV_m}{s_{11}+s_{12}} + \frac{c\,V_n(B_{11}-B_{12})}{s_{11}-s_{12}}\right)}, \quad (11a)$$

$$k_3 = \pm\sqrt{-\frac{1}{2w}\left(g - \frac{B_{12}(2u_m - V_m c)}{s_{11}+s_{12}}\right)}. \quad (11b)$$

It is instructive to rewrite and analyze expressions (11) in a dimensionless form, namely $\frac{k_1}{k_0} = \pm\sqrt{-\frac{g}{|g|} - \frac{g_a}{|g|} + b_1 u}$ and $\frac{k_3}{k_0} = \pm\sqrt{-\frac{g}{|g|} + b_3 u}$, where $k_0 = \sqrt{\frac{|g|}{2w}}$ is a characteristic wave number, $u = u_m - cV_m/2$ is a dimensionless "effective" strain; $b_1 = \frac{B_{11}+B_{12}}{|g|(s_{11}+s_{12})}$ and $b_3 = \frac{2B_{12}}{|g|(s_{11}+s_{12})}$ are dimensionless gradient-related parameters. Below we will analyze the positive roots $k_{1,3} > 0$ only, since the negative ones describe the same physical states. Here we also introduced anisotropy term $g_a = (B_{11} - B_{12})\frac{c\,V_n}{2(s_{11}-s_{12})}$.

**Figure 2** illustrates the square-root like dependences of the dimensionless wave-vector components, $k_1/k_0$ and $k_3/k_0$, on the effective strain $u$ for $g > 0$ (red and magenta curves) and $g < 0$ (black and blue curves). Note that positive $w$ determines the existence of the modulation and its characteristic wave number.

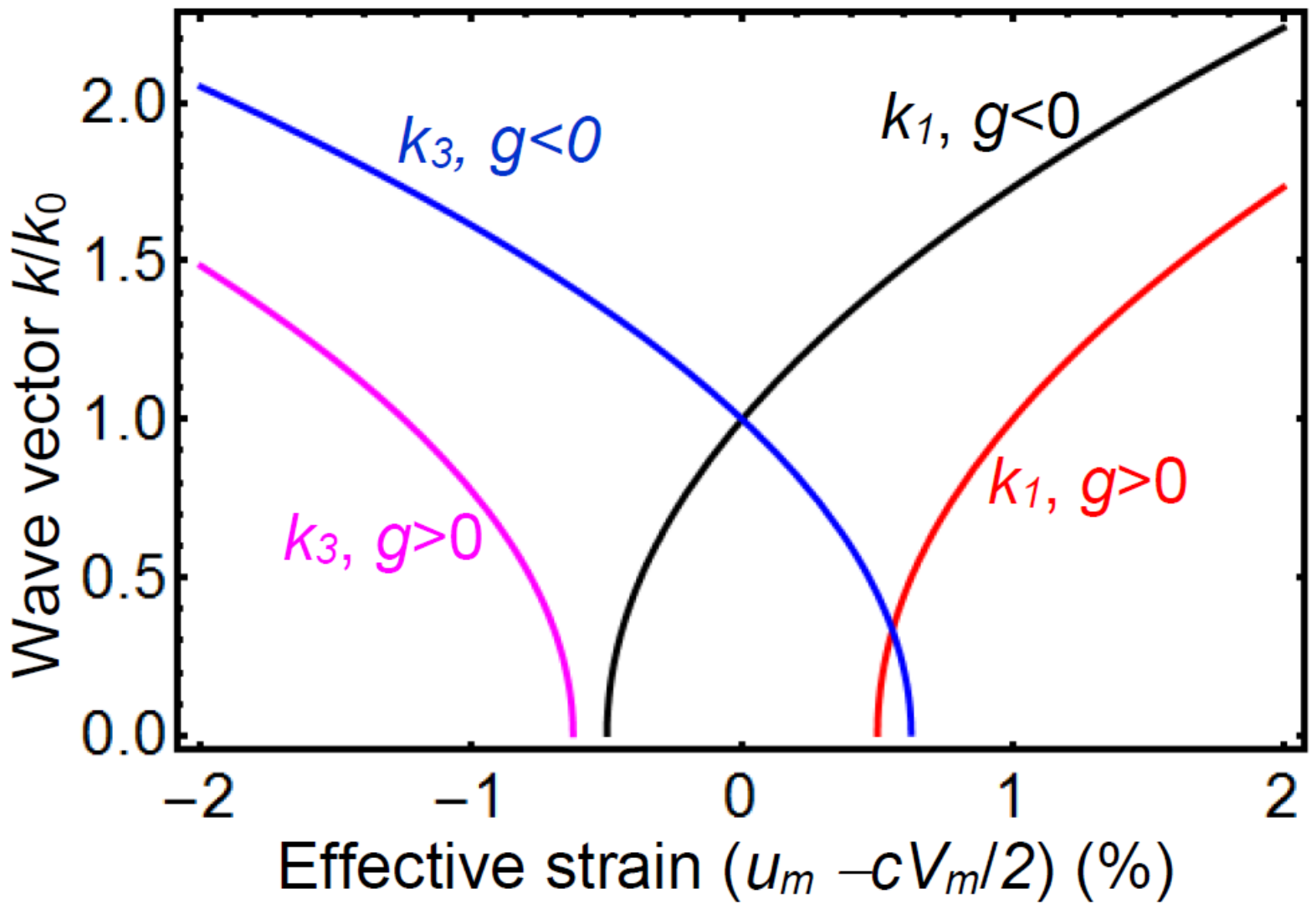

**FIGURE 2.** The dependence of the wave-vector component $k_1/k_0$ and $k_3/k_0$ on the effective strain $(u_m - cV_m/2)$ calculated from Eqs. (11) for $g > 0$ (red and magenta curves) and $g < 0$ (black and blue curves), parameters $b_1 = 200$ and $b_3 = -160$.

The transition temperatures, corresponding to the $\eta$-modulation vectors given by Eq. (11), are

$$T_{pp}(u_m, c) = -\frac{c}{k_B}\left[\alpha - \frac{1}{4w}\left(2g - (B_{11}+B_{12})\frac{2u_m - cV_m}{s_{11}+s_{12}} + (B_{11}-B_{12})\frac{c\,V_n}{s_{11}-s_{12}}\right)^2\right] \quad (12a)$$

$$T_{pr}(u_m, c) = -\frac{c}{k_B}\left[\alpha - \frac{1}{2w}\left(g - B_{12}\frac{2u_m - V_m c}{s_{11}+s_{12}}\right)^2\right] \quad (12b)$$

The conditions of the equilibrium between the ordered phases can be obtained from the equality of the corresponding free energy densities, Eqs. (8c) and (9c), namely

$$\frac{(\beta_{pr}c + k_B T)^2}{\gamma_{pr}c^3 + k_B T} = \frac{(\beta_{pp}c + k_B T)^2}{\gamma_{pp}c^3 + k_B T}, \quad (13a)$$

allowing for the conditions of the long-range orders existence

$$\beta_{pr}c + k_B T < 0, \qquad \beta_{pp}c + k_B T < 0. \quad (13b)$$

Note that the expressions (11) for **k**-vector components should be substituted in Eqs. (13).

Equation (13a) can be solved with respect to temperature in the following form:

$$T = \frac{c}{k_B}\frac{({\beta_{pr}}^2 - {\beta_{pp}}^2) + 2c^2(\beta_{pr}\gamma_{pp} - \beta_{pp}\gamma_{pr}) \pm \mathrm{Det}[\beta_{pp}, \beta_{pr}]}{4(\beta_{pp} - \beta_{pr}) - 2(\gamma_{pp} - \gamma_{pr})c^2}, \quad (14a)$$

$$\mathrm{Det}[\beta_{pp}, \beta_{pr}] = \sqrt{(\beta_{pp} - \beta_{pr})^4 + 4(\beta_{pp} - \gamma_{pp}c^2)(\beta_{pr} - \gamma_{pr}c^2)(\beta_{pp} - \beta_{pr})^2}. \quad (14b)$$

The signs "±" in Eq. (14a) correspond to two different roots, which have physical sense if correspond to $T\geq 0$. These roots should further satisfy two conditions (13b). Depending on the parameters it appeared possible at least for the largest root of Eq. (14a). Since the sign of denominator in Eq. (14a) is not fixed, both signs in the numerator, $+\mathrm{Det}[\beta_{pp},\beta_{pr}]$ or $-\mathrm{Det}[\beta_{pp},\beta_{pr}]$, are possible depending on the parameters. Hence, the number and selection of the roots in Eq. (14a) should be established numerically depending on the number and values of the fitting parameters.

It can be shown that, at the chosen parameter values, one of the roots (14a) is close to the expression $\left(T_{pp}(u_m,c)+T_{pr}(u_m,c)\right)/2$. Hence, for $\beta_{pp}\approx\beta_{pr}$ expression (14a) can be expanded in series on the small difference $(\beta_{pp}-\beta_{pr})$ powers, namely

$$T\approx\frac{c}{k_B}\left(\beta_{pp}-\beta_{pr}\right)\left(-\frac{1}{2}+\frac{\gamma_{pp}c^2-\beta_{pp}\pm\sqrt{(\gamma_{pp}c^2-\beta_{pp})(\gamma_{pr}c^2-\beta_{pr})}}{(\gamma_{pp}-\gamma_{pr})c^2}\right). \quad (14c)$$

When a strong inequality $|\gamma_{pp}-\gamma_{pr}|<<\gamma_{pp}$ takes place, the second term in parenthesis in Eq. (14c) could be either close to unity (sign "–") or much higher than unity (sign "+"). It is seen that in the case of sign "–" the solutions (14b)-(14c) will not satisfy both of relations (13b) simultaneously. The other root in Eq. (14a) deviates more significantly from the value $[T_{pp}(u_m,c)+T_{pr}(u_m,c)]/2$, thus, in this case, conditions (13b) will be satisfied.

Using Eqs. (10)-(14) we plotted phase diagrams of the system and corresponding modulation amplitudes and wave vectors, shown in **Fig. 3-4** and **Fig. 5**, respectively**.** Note that the ordered phases $OP_{\perp}$ and $OP_{||}$ have different orientation of the order parameter modulation amplitude, $\eta(x)$ and $\eta(z)$, with respect to the film surfaces.

**Fig. 3(a)** represents the scheme of the $\eta(x,z)$ distribution in the ordered ($OP_{\perp}$ and $OP_{||}$) and disordered (DP) phases. The ordered phases $OP_{\perp}$ and $OP_{||}$ are characterized by periodic changes of $\eta=\frac{c}{c_0}(\delta n_b-\delta n_a)$, which are proportional to the difference of the sublattices "a" and "b" occupation numbers, $\delta n_b-\delta n_a$. The total concentration "c" remains constant, since the sum $\delta n_a+\delta n_b$ is independent on $\eta$. The $OP_{||}$ phase is characterized by a periodic change $\eta\sim|\tilde{\eta}|\cos[k_3x_3+\delta]$, corresponding to the long-range modulation of the defect sub-lattices "a" and "b" occupation perpendicular to the substrate plane $x_3=h$. The $OP_{\perp}$ phase is characterized by a periodic change $\eta\sim|\tilde{\eta}|\cos[k_1x_1+\delta]$, corresponding to the long-range modulation of the defect sub-lattices "a" and "b" occupation parallel to the substrate plane $x_3=h$. The disordered DP phase is characterized by $\eta = 0$, corresponding to the equal filling of both sub-lattices. Note that the modulation periods, $\frac{2\pi}{k_1}$ and $\frac{2\pi}{k_3}$, should be significantly larger than the lattice constant for the validity of the continuum approach, and so **Fig. 3a** illustrates only these long-range modulations of the order parameter $\eta(\boldsymbol{x})$, but not the distance between the defect sub-lattices planes.

**Figures 3b-e** and **4a-d** are phase diagrams of an ultra-thin film in dependence on the misfit strain $u_m$ and the dimensionless defect concentration $c/c_0$calculated for negative and positive α, respectively. Corresponding modulation wave vectors, $k_{\perp}$ and $k_{||}$, are shown in **Figs. 5a** and **5b**, respectively.

**Figures 3(b-e),** calculated for negative α, demonstrate the presence of $OP_{\perp}$, $OP_{||}$ and DP phases, whose boundaries depend on the gradient $g$ and striction $B_{ij}$ coefficients, which are different for **Figs. 3(b-e)**. The ultrathin almost vertical dark green stripe between $OP_{||}$ and $OP_{\perp}$ phases located at small $|u_m| < 0.1\%$ in **Fig. 3b** and **3c** is the coexistence region of the ordered phases, where the wave vector becomes very small, $k \to 0$. This phase is absent for the diagrams in **Figs. 3d** and **3e**, which have a tricritical point, where $OP_{\perp}$, $OP_{||}$ and DP phases coexist. The tricritical point {$u_m$=0, $c/c_0$ =0.63} is almost independent on $g$ and $B_{ij}$ signs. The $OP_{||}$-DP and $OP_{\perp}$-DP boundaries are curved for all diagrams. DP phase occupies mountain-shape region with the "top" at {$u_m$=0, $c/c_0$ =0.63}, which hill-sides start at small misfits. The region of its stability gradually decreases with $|u_m|$ increase. At that the noticeable asymmetry between compressive ($u_m < 0$) and tensile ($u_m > 0$) misfit strains is evident and related with positive Vegard effect ($V_m > 0$). It can be deduced from **Fig. 3(b-e)**, that the change of the $B_{ij}$ sign leads to an interchange between the $OP_{||}$ and $OP_{\perp}$ phases with different orientation of η-ordering, stemming from the chemo-strictive coupling between the strain $u_m$ and defect concentration $c$, expressed by the coupling terms $\frac{B_{11}+B_{12}}{s_{11}+s_{12}}(u_m - cV_m/2)$ and $\frac{2B_{12}}{s_{11}+s_{12}}(u_m - cV_m/2)$.

The $OP_{||}$-$OP_{\perp}$ boundary is almost vertical, and its small slope weakly depends on the $V_m$ value. The $OP_{\perp}$ phase exists for compressive misfit strain $u_m < 0$ at $B_{11} + B_{12} < 0$, and for tensile misfit strain $u_m > 0$ at $B_{11} + B_{12} > 0$. The $OP_{||}$ phase exists for compressive misfit strain $u_m < 0$ at $B_{11} + B_{12} > 0$, and for tensile misfit strain $u_m > 0$ at $B_{11} + B_{12} < 0$. The color maps of the order parameter amplitudes [$\eta(x_1)$ and $\eta(x_3)$] and wave vectors ($k_{\perp}$ and $k_{||}$) calculated for $\alpha < 0$, $g > 0$, $B_{11} + B_{12} > 0$ are shown in **Figs. 5a** and **5c**, respectively**.** The continuous transition from $OP_{\|}$ to $OP_{\perp}$ phase occurs at $u_m = 0$. Notably that $k_{\perp}=k_{\|}=0$ in the region $u_m \approx 0$. The values of $\eta$ and $k$ gradually increase with $|u_m|$ increase.

The change of $\alpha$ sign critically affects the phase diagrams, as is shown in **Fig. 4**. The most pronounced effect is the appearance of the wide DP region between the ordered $OP_{\perp}$ and $OP_{\|}$ phases, so that the tricritical point is absent for all concentrations "$c$". Note that the DP exists in the regions of misfits $|u_m| \leq 2\%$ and this region grows with defect concentration decrease. The width of the DP region slightly increases with $|u_m|$ increase. The change of the $B_{ij}$ sign leads to the interchange between the $OP_{\|}$ and $OP_{\perp}$ phases, while the effect of the coefficient g is negligibly small in this case [compare **Figs. 4a-b** with **4c-d**]. The $OP_{\perp}$ phase exists for high compressive misfit strain $u_m < -2.5\%$ at $B_{11} + B_{12} < 0$, and for tensile misfit strain $u_m > +2.5\%$ at $B_{11} + B_{12} > 0$. The $OP_{\|}$ phase exists for compressive misfit strain $u_m < -1.9\%$ at $B_{11} + B_{12} > 0$, and for tensile misfit strain $u_m > +1.9\%$ at $B_{11} + B_{12} < 0$. The color maps of the order parameter amplitudes [$\eta(x_1)$ and $\eta(x_3)$] and wave vectors ($k_{\perp}$ and $k_{\|}$) calculated for $\alpha > 0$, $g > 0$, $B_{11} + B_{12} > 0$ are shown in **Figs. 5b** and **5d**, respectively**.** The $OP_{\|}$ to $OP_{\perp}$ phases are separated by a wide region of the DP phase located at $|u_m| \leq 2\%$. The values of k are nonzero at the DP-$OP_{\|}$ and DP-$OP_{\perp}$ boundaries. The values of $\eta$ and $k$ gradually increase with $|u_m|$ increase.

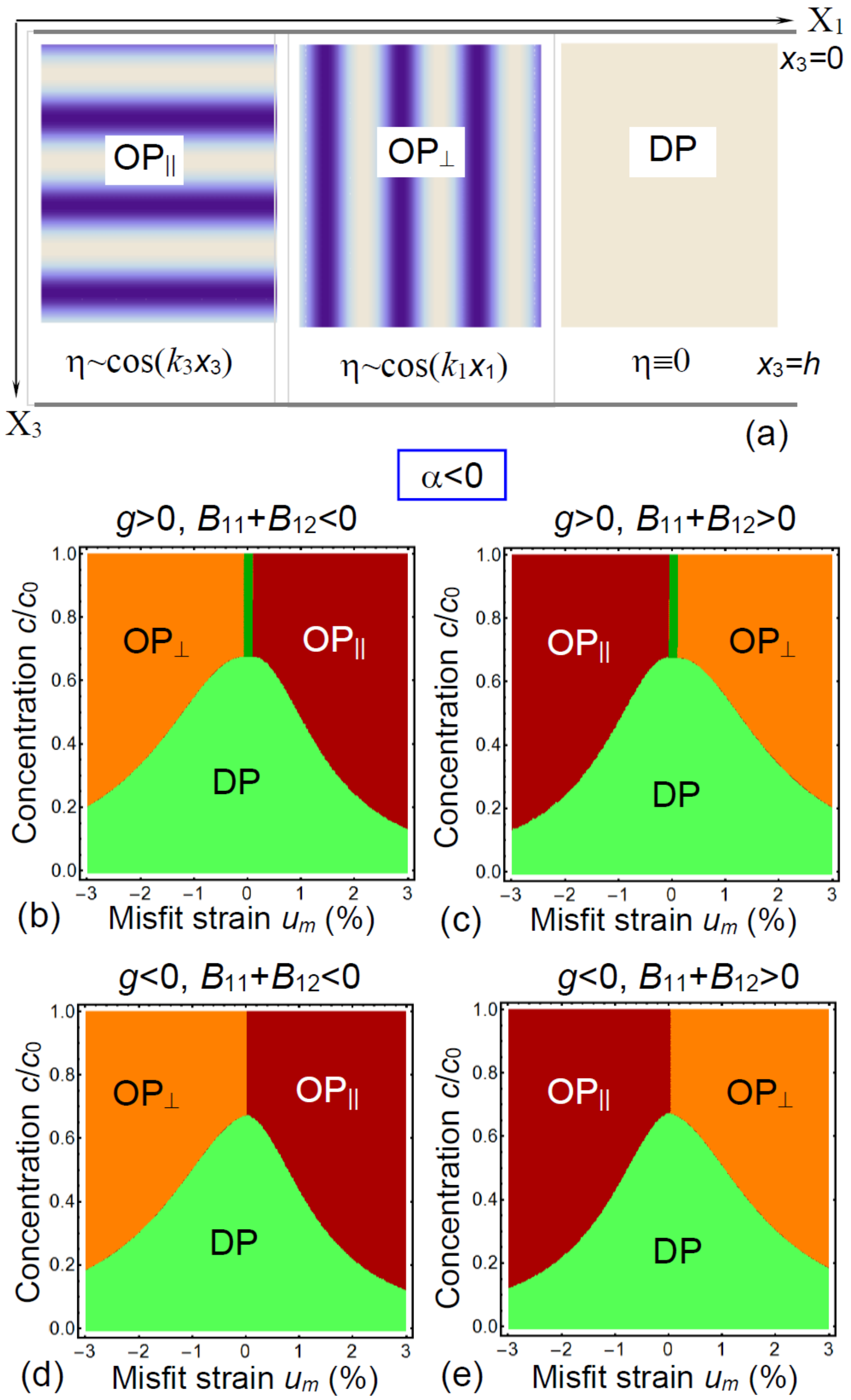

**FIGURE 3. (a)** Geometry of the problem and schematics of different phases characterized by different directions of the order parameter $\eta$ modulation . The $OP_{||}$ phase is characterized by a periodic change $\eta \sim |\tilde{\eta}| \cos[k_3 x_3 + \delta]$ perpendicular to the substrate plane $x_3 = h$. The $OP_{\perp}$ phase is characterized by a periodic change $\eta \sim |\tilde{\eta}| \cos[k_1 x_1 + \delta]$ parallel to the substrate plane $x_3 = h$. The disordered DP phase is characterized by $\eta = 0$. **(b)-(e)** Phase diagrams in dependence on the normalized defect concentration $c/c_0$ and misfit strain $u_m$ calculated for negative α, different signs of coefficient $g$ and striction coefficients $B_{ij}$

(indicated above the plots) at room temperature T=293 K. Parameters: $\alpha c_0^2 = -6\times10^4$ J·m$^{-3}$, $|g|c_0^2 = 10^{-17}$ J/m, $wc_0^2 = 3\times10^{-37}$ J·m, $s_{11} = 4\times10^{-12}$ Pa$^{-1}$, $s_{12} = -1\times10^{-12}$ Pa$^{-1}$, $V_m = 30$ Å$^3$, while striction coefficients $B_{11}c_0^2 = 5\times10^{-26}$ J·m$^2$, and $B_{12}c_0^2 = -2\times10^{-26}$ J·m$^2$, and maximal (steric limit) concentration of defects is $c_0$=$10^{25}$ m$^{-3}$.

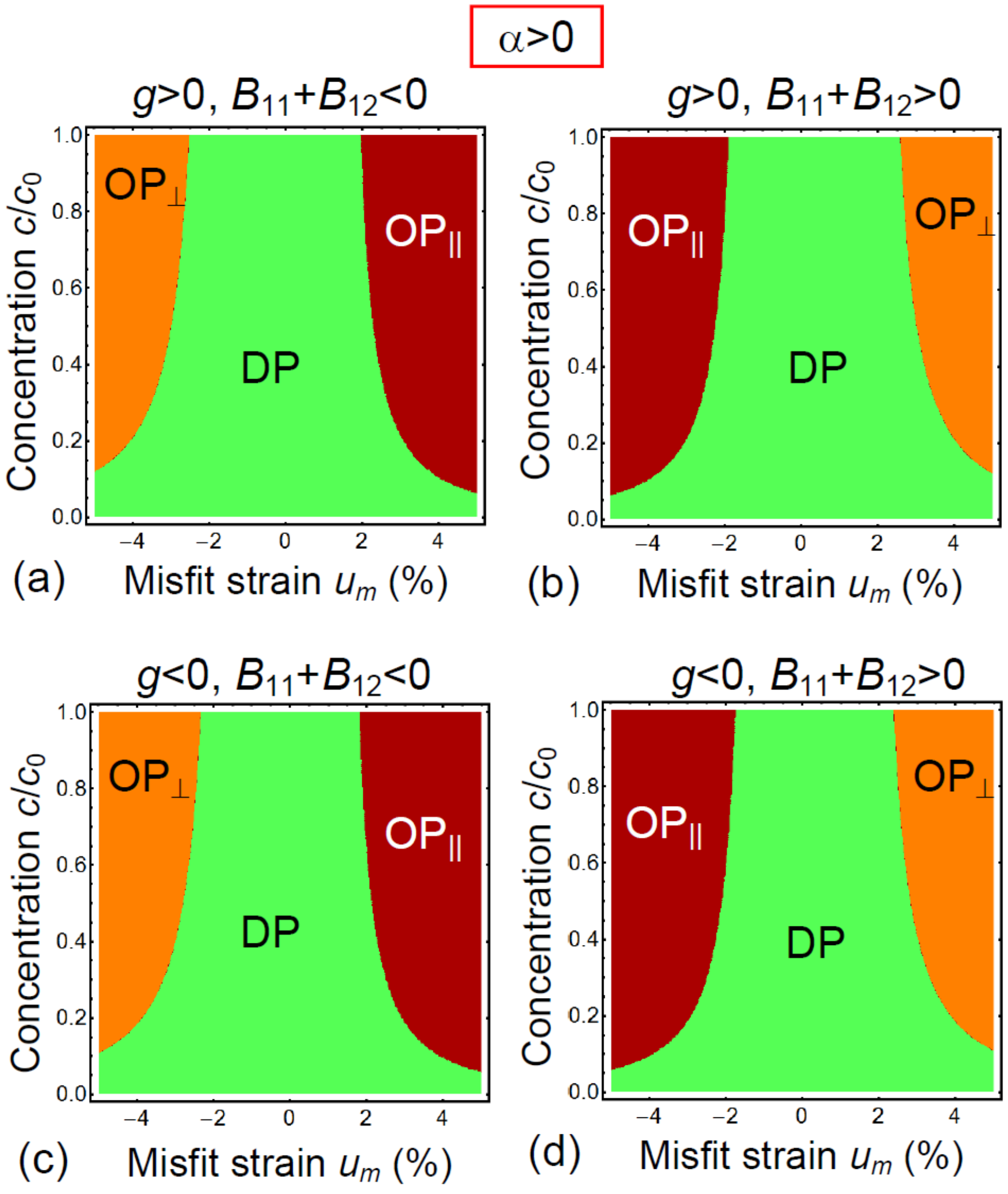

**FIGURE 4.** Phase diagrams in dependence on normalized defect concentration $c/c_0$ and misfit strain $u_m$ calculated for positive α, different signs of gradient coefficient $g$ and striction coefficients $B_{ij}$ (indicated above the plots), and $\alpha > 0$. Other parameters are the same as in **Fig. 3**.

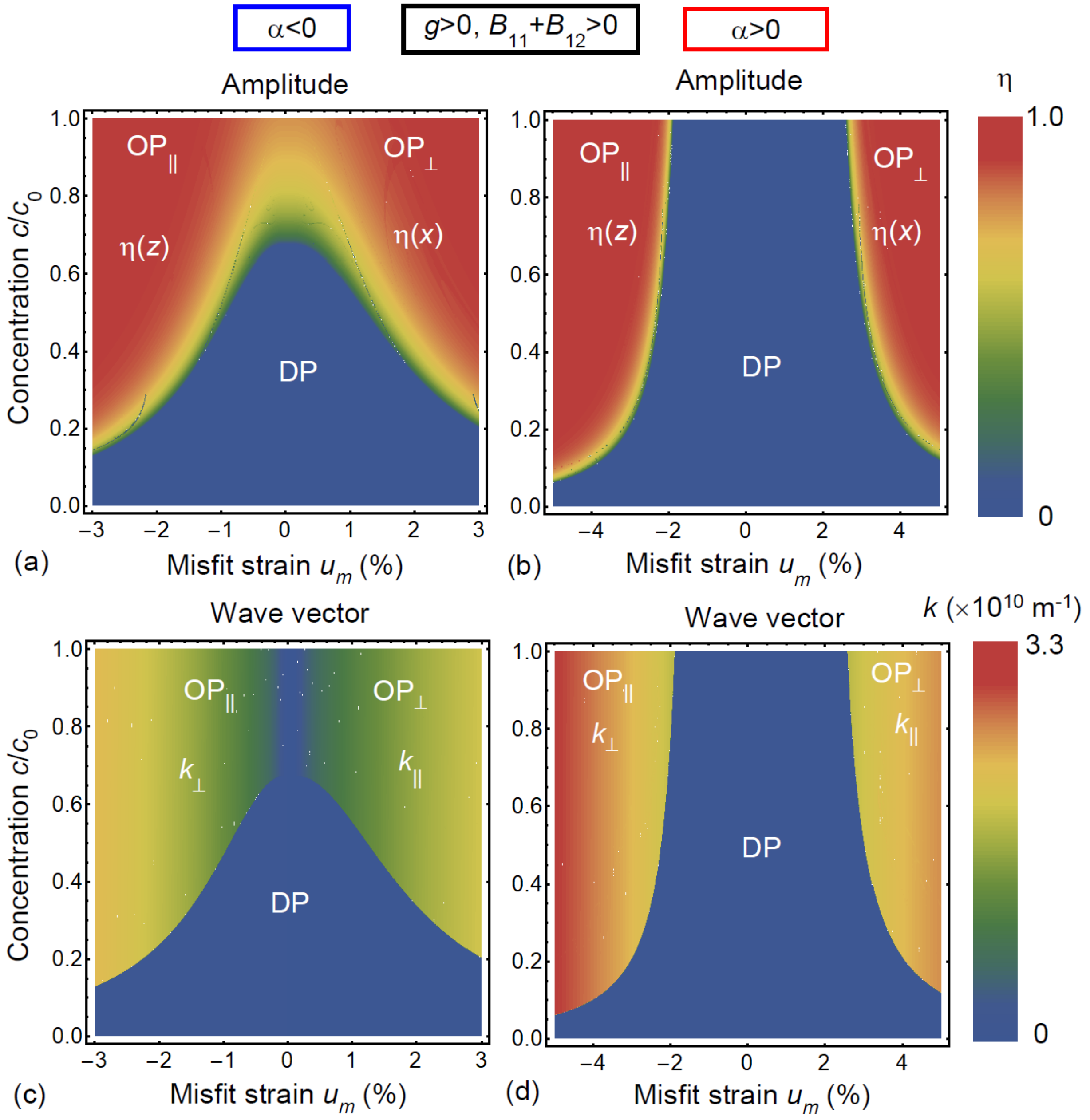

**FIGURE 5.** Order parameter amplitudes $\eta(x_1)$ and $\eta(x_3)$ **(a, b)** and wave vectors **(c, d)** in dependence on normalized defect concentration $c/c_0$ and misfit strain $u_m$ calculated for negative α **(a, c)** and positive α **(b, d)**; $g > 0$, and $B_{11} + B_{12} > 0$. Numerical values of parameters are the same as in **Fig. 3**.

Note that the values of the order parameter η and the components of the wave vector, $k_1$ and $k_3$, were calculated on the basis of numerical minimization of the more complex free energy functional [see Eqs. (8a) and (9a)], for which the series expansion was not used. It turned out that the differences between the approximate expressions (11) and the results of numerical calculations are insignificant for the wave vector. Also, to describe the phase transitions between the ordered and disordered phases, expressions (12) exactly correspond to numerical calculations, since these transitions are the second-order transitions and, the order parameter can be considered small near the transition points.

Note that phenomenological parameters listed in the caption to **Fig. 3** were selected in such a way so as to satisfy the physical conditions $\alpha c_0 \cong k_B T_{room}$, $\sqrt{|\alpha|/g} \geq k_0$, $\sqrt{|g|/w} < k_0$, while $s_{ij}$ and $c_0$ values are typical for oxides. As for the striction coefficients $B_{ij}$, these are chosen so that the combinations of parameters, $\frac{B_{11}+B_{12}}{s_{11}+s_{12}}(u_m - cV_m/2)$ and $\frac{2B_{12}}{s_{11}+s_{12}}(u_m - cV_m/2)$, are of the same order as $|g|$. To summarize the section, results shown in **Fig. 3-5** indicate that one can control the defect ordering-disordering by changing their concentration $c/c_0$ and misfit strain $u_m$ at fixed values of the other parameters.

Finally it may be interesting to compare the defect ordering with the cycloid ordering in a strained multiferroic $BiFeO_3$ thin films (see e.g. Ref. [21-23] and refs. to original papers therein). Actually, the strain-induced singularity of the cycloid period and phase diagram with various spin cycloid orientations in $BiFeO_3$ can be topologically analogous to the one considered in **Fig. 3-5** proving that chiral incommensurate phases of defect ordering (allowed by the theoretical formalism) can exist in the strained film.

## V. CONCLUSION

We have analyzed the ordering of defects (e.g. oxygen vacancies) in thin oxide films in the framework of the continuum Landau-type theory. We derived analytical expressions for the energies of various defect-ordered states and calculated and analyzed phase diagrams dependence on the film-substrate misfit strain and concentration of defects for different gradient, striction and Vegard coefficients.

We have found that two defect-ordered phases, which are characterized by either parallel or perpendicular defect ordering in planes and corresponding wave vectors, can be stable. The stability conditions are determined by the misfit strain and the defect concentration at fixed values of the other parameters. The ordered phases border with the defect-disordered phase. Hence, we have shown that it is possible to control the defect ordering-disordering by changing their concentration and the film-substrate misfit strain (compressive or tensile). Thus, the obtained results open possibilities to create and control superstructures of ordered defects in thin oxide films by selecting the appropriate substrate and defect concentration.

**Author contributions.** A.N.M. and Y.A.G. generated the research idea, stated the problem, and performed analytical calculations jointly with E.A.E. E.A.E. wrote the codes and performed numerical calculations. A.N.M., S.V.K. and Y.A.G. wrote the manuscript draft. All co-authors worked intensively on the results interpretation and manuscript improvement.

**Acknowledgements.** This material is based upon work (S.V.K.) supported by the U.S. Department of Energy, Office of Science, Office of Basic Energy Sciences, and performed in the Center for Nanophase Materials Sciences, supported by the Division of Scientific User Facilities. This work was supported by the Deutsche Forschungsgemeinschaft (DFG) via the grant No. 405631895 (GE-1171/8-1). A portion of FEM was conducted at the Center for Nanophase Materials Sciences, which is a DOE Office of Science User Facility. M.V.S. also acknowledges Russian academic excellence project "5-100" for Sechenov University and the Ministry of Science and Higher Education of the Russian Federation State assignment 2020– 2022 No. FSMR-2020-0018 (Proposal mnemonic code 0719-2020-0018). This project (A.N.M.) has received funding from the European Union’s Horizon 2020 research and innovation programme under the Marie Skłodowska-Curie grant agreement No 778070 – TransFerr – H2020-MSCA-RISE-2017.

## SUPPLEMENTARY MATERIALS

**to**

## MESOSCOPIC THEORY OF DEFECT ORDERING-DISORDERING TRANSITIONS IN THIN OXIDE FILMS

### APPENDIX. Derivation of the Free Energy Expression

Let us denote the long-range modulation of the order parameter $\eta$, and wave vector of modulation as $k_i$,where $i$=1, 2, 3 denotes different components of vector, determining the orientations of ordering (modulation) planes, so that we suppose the following distribution $\eta$:

$$\eta = \eta_0 + \tilde{\eta}\ \exp[i(k_1x_1 + k_2x_2 + k_3x_3)] + c.c. \qquad \text{(A.1)}$$

Correspondingly, the gradient of $\eta$ is

$$\frac{\partial\eta}{\partial x_j} = ik_j\ \tilde{\eta}\ \exp[i(k_1x_1 + k_2x_2 + k_3x_3)] + c.c. \qquad \text{(A.2)}$$

One could easily find from (A.1) and (A.2) the mean square average of the $\eta$ gradient

$$\langle\frac{\partial\eta}{\partial x_i}\frac{\partial\eta}{\partial x_j}\rangle \cong 2|\ \tilde{\eta}|^2 k_i k_j \qquad \text{(A.3)}$$

Note, that harmonic functions disappeared after the averaging.

General expression for order parameter gradient contribution the free energy density is

$$\frac{g_{ij}}{2}\frac{\partial\eta}{\partial x_i}\frac{\partial\eta}{\partial x_j} + B_{ijkl}\frac{\sigma_{ij}}{2}\left(\frac{\partial\eta}{\partial x_k}\frac{\partial\eta}{\partial x_l} + \frac{\partial\eta}{\partial x_l}\frac{\partial\eta}{\partial x_k}\right) \qquad \text{(A.4)}$$

Using expressions (A.1)-(A.4) different contributions to the free energy of the system could be written as follows:

$$\langle g_{ij}\frac{\partial\eta}{\partial x_i}\frac{\partial\eta}{\partial x_j}\rangle \cong 2g_{ij}k_ik_j|\ \tilde{\eta}|^2 \qquad \text{(A.5)}$$

$$\langle \mathrm{B}_{ijkl}\sigma_{ij}\frac{\partial\eta}{\partial x_k}\frac{\partial\eta}{\partial x_l}\rangle = 2\mathrm{B}_{ijkl}k_ik_j|\ \tilde{\eta}|^2\sigma_{ij}, \qquad \text{(A.6)}$$

$$\langle\eta\rangle = \langle\eta_0 + 2|\tilde{\eta}|\cos[(\boldsymbol{kx} + \delta)]\rangle \cong \eta_0 \qquad (A.7)$$

Taking into account (A.5)-(A.6) and using isotropic approximation, $g_{ij} = g\delta_{ij}$, the gradient and elastic contributions of free energy can be expanded on the powers of $k_i$. Assuming that the anisotropic Vegard tensor is diagonal (or at least can be diagonalized), that is true for many cases [55], result has the following form

$$\Delta F_{FE} = gc^2(k_1^2 + k_2^2 + k_3^2)|\tilde{\eta}|^2 - B_{11}c^2(\sigma_{11}k_1^2 + \sigma_{22}k_2^2 + \sigma_{33}k_3^2)|\tilde{\eta}|^2$$
$$- B_{12}c^2(\sigma_{11}k_2^2 + \sigma_{22}k_1^2 + \sigma_{11}k_3^2 + \sigma_{33}k_1^2 + \sigma_{22}k_3^2 + \sigma_{33}k_2^2)|\tilde{\eta}|^2$$
$$- B_{44}c^2(\sigma_{12}k_1k_2 + \sigma_{13}k_1k_3 + \sigma_{23}k_2k_3)|\tilde{\eta}|^2 - \frac{s_{11}}{2}(\sigma_{11}^2 + \sigma_{22}^2 + \sigma_{33}^2)$$
$$- s_{12}(\sigma_{11}\sigma_{22} + \sigma_{11}\sigma_{33} + \sigma_{22}\sigma_{33}) - \frac{s_{44}}{2}(\sigma_{12}^2 + \sigma_{13}^2 + \sigma_{23}^2) - c\left(V_{11}^c + \eta_0 V_{11}^\eta\right)\sigma_{11}$$
$$- c\left(V_{22}^c + \eta_0 V_{22}^\eta\right)\sigma_{22} - c\left(V_{33}^c + \eta_0 V_{33}^\eta\right)\sigma_{33} \quad (A.7)$$

Voight matrix notations is used in Eq.(A.7), and $B_{ij}$ are the components of the fourth rank tensor in these notations, which are different from the second rank tensor, $\tilde{B}_{ij}$, introduced in the main text. Voight notations are

$$s_{1111} = s_{11},\ \ s_{1122} = s_{12},\ \ 4s_{1212} = s_{44}, \quad (A.8)$$

$$B_{1111} = B_{11},\ \ B_{1122} = B_{12},\ \ 4B_{1212} = B_{44}, \quad (A.9)$$

Modified Hooke's law could be obtained from the relation $u_{ij} = -\partial(\Delta F_{FE})/\partial\,\sigma_{ij}$:

$$u_{11} = s_{11}\sigma_{11} + s_{12}\sigma_{22} + s_{12}\sigma_{33} + \tilde{B}_{11}|\tilde{\eta}|^2c^2 + c\left(V_{11}^c + \eta_0 V_{11}^\eta\right), \quad (A.10)$$

$$u_{22} = s_{12}\sigma_{11} + s_{11}\sigma_{22} + s_{12}\sigma_{33} + \tilde{B}_{22}|\tilde{\eta}|^2c^2 + c\left(V_{22}^c + \eta_0 V_{22}^\eta\right), \quad (A.11)$$

$$u_{33} = s_{12}\sigma_{11} + s_{12}\sigma_{22} + s_{11}\sigma_{33} + \tilde{B}_{33}|\tilde{\eta}|^2c^2 + c\left(V_{33}^c + \eta_0 V_{33}^\eta\right), \quad (A.12)$$

$$u_{23} = s_{44}\sigma_{23} + \tilde{B}_{23}|\tilde{\eta}|^2c^2,\ \ u_{13} = s_{44}\sigma_{13} + \tilde{B}_{13}|\tilde{\eta}|^2c^2, u_{12} = s_{44}\sigma_{12} + \tilde{B}_{12}|\tilde{\eta}|^2c^2. \quad (A.13)$$

Here we used the designation for the convolution with wave vector:

$$\tilde{B}_{ij} \equiv B_{ijmn}k_mk_n$$

For instance, one has the following relations: $\tilde{B}_{11} \equiv B_{11}k_1^2 + B_{12}(k_2^2 + k_3^2)$, $\tilde{B}_{22} \equiv B_{11}k_2^2 + B_{12}(k_1^2 + k_3^2)$, $\tilde{B}_{33} \equiv B_{11}k_3^2 + B_{12}(k_1^2 + k_2^2)$, $\tilde{B}_{12} \equiv B_{44}k_1\,k_2$ , $\tilde{B}_{13} \equiv B_{44}k_1\,k_3$ , and $\tilde{B}_{23} \equiv B_{44}k_2\,k_3$ .

The solution for the misfit of thin film with its substrate is well known. For the film with normal along $X_3$ one has the following relations for some of stress and strain components:

$$\sigma_{13} = \sigma_{23} = \sigma_{33} = 0 \quad (A.14)$$

$$u_{11} = u_m, \quad u_{22} = u_m, \ \ u_{12} = 0 \quad (A.15)$$

Here $u_m$ is the misfit strain. For the sake of simplicity let us consider the case of only one zero component, $k_2 = 0$. Therefore, taking (A.14) and (A.15) into account, one could rewrite (A.10)-(A.13) in the following form:

$$u_m = s_{11}\sigma_{11} + s_{12}\sigma_{22} + \tilde{B}_{11}|\tilde{\eta}|^2c^2 + c\left(V_{11}^c + \eta_0 V_{11}^\eta\right), \quad (A.16a)$$

$$u_m = s_{12}\sigma_{11} + s_{11}\sigma_{22} + \tilde{B}_{22}|\tilde{\eta}|^2c^2 + c\left(V_{22}^c + \eta_0 V_{22}^\eta\right), \quad (A.16b)$$

$$u_{33} = s_{12}\sigma_{11} + s_{12}\sigma_{22} + \tilde{B}_{33}|\,\tilde{\eta}|^2 c^2 + c\left(V_{33}^{c} + \eta_0 V_{33}^{\eta}\right), \quad \text{(A.16c)}$$

$$u_{23} = \tilde{B}_{23}|\tilde{\eta}|^2 c^2,\; u_{13} = \tilde{B}_{13}|\,\tilde{\eta}|^2 c^2, 0 = s_{44}\sigma_{12} + \tilde{B}_{12}|\,\tilde{\eta}|^2 c^2. \quad \text{(A.16d)}$$

The solution of the system (A.16) is

$$\sigma_{11} = \frac{2u_m - (V_{11}^{c} + \eta_0 V_{11}^{\eta})c - (V_{22}^{c} + \eta_0 V_{22}^{\eta})c - (\tilde{B}_{11} + \tilde{B}_{22})|\,\tilde{\eta}|^2 c^2}{2(s_{11} + s_{12})} - \frac{(V_{11}^{c} + \eta_0 V_{11}^{\eta})c - (V_{22}^{c} + \eta_0 V_{22}^{\eta})c + (\tilde{B}_{11} - \tilde{B}_{22})|\,\tilde{\eta}|^2 c^2}{2(s_{11} - s_{12})}$$

$$\sigma_{22} = \frac{2u_m - (V_{11}^{c} + \eta_0 V_{11}^{\eta})c - (V_{22}^{c} + \eta_0 V_{22}^{\eta})c - (\tilde{B}_{11} + \tilde{B}_{22})|\,\tilde{\eta}|^2 c^2}{2(s_{11} + s_{12})} + \frac{(V_{11}^{c} + \eta_0 V_{11}^{\eta})c - (V_{22}^{c} + \eta_0 V_{22}^{\eta})c + (\tilde{B}_{11} - \tilde{B}_{22})|\,\tilde{\eta}|^2 c^2}{2(s_{11} - s_{12})}$$

$$u_{33} = \left(V_{33}^{c} + \eta_0 V_{33}^{\eta}\right)c + \tilde{B}_{33}|\,\tilde{\eta}|^2 c^2 +$$

$$+\frac{s_{12}}{s_{11} + s_{12}}\left[2u_m - \left(V_{11}^{c} + \eta_0 V_{11}^{\eta} + V_{22}^{c} + \eta_0 V_{22}^{\eta}\right)c - \left(\tilde{B}_{11} + \tilde{B}_{22}\right)c^2|\,\tilde{\eta}|^2\right]$$

$$\sigma_{12} = -\frac{\tilde{B}_{12}|\,\tilde{\eta}|^2 c^2}{s_{44}}, u_4 = \tilde{B}_{23}|\tilde{\eta}|^2 c^2, u_5 = B_{44}k_1 k_3|\,\tilde{\eta}|^2 c^2. \quad \text{(A.17)}$$

Finally, free energy renormalization could be obtained from the Legendre transformation of the initial free energy (A.7), $\Delta\tilde{F}_{FE} = \Delta F_{FE} + \sigma_{11}u_{11} + \sigma_{22}u_{22}$ , that is:

$$\Delta\tilde{F}_{FE} = gc^2(k_1^2 + k_2^2 + k_3^2)|\,\tilde{\eta}|^2 - \left(\tilde{B}_{11}\sigma_{11} + \tilde{B}_{12}\sigma_{12} + \tilde{B}_{22}\sigma_{22} + \tilde{B}_{33}\sigma_{33}\right)|\,\tilde{\eta}|^2 c^2 - c\left(\left(V_{11}^{c} + \eta_0 V_{11}^{\eta}\right)\sigma_{11} + \left(V_{22}^{c} + \eta_0 V_{22}^{\eta}\right)\sigma_{22}\right) - \frac{s_{11}}{2}(\sigma_{11}^2 + \sigma_{22}^2) - s_{12}(\sigma_{11}\sigma_{22}) - \frac{s_{44}}{2}\sigma_{12}^2 + \sigma_{11}u_m + \sigma_{22}u_m \equiv gc^2(k_1^2 + k_2^2 + k_3^2)|\,\tilde{\eta}|^2 + \frac{\left(\tilde{B}_{12}|\,\tilde{\eta}|^2 c^2\right)^2}{2s_{44}} + \frac{((V_{11}^{\eta} - V_{22}^{\eta})\eta_0 c + (V_{11}^{c} - V_{22}^{c})c + (\tilde{B}_{11} - \tilde{B}_{22})|\,\tilde{\eta}|^2 c^2)^2}{4(s_{11} - s_{12})} + \frac{((V_{11}^{\eta} + V_{22}^{\eta})\eta_0 c + (V_{11}^{c} + V_{22}^{c})c + (\tilde{B}_{11} + \tilde{B}_{22})|\tilde{\eta}|^2 c^2 - 2u_m)^2}{4(s_{11} + s_{12})} \quad \text{(A.18b)}$$

Expansion on the powers of η could be

$$\Delta\tilde{F}_{FE} = \left[g(k_1^2 + k_2^2 + k_3^2) + \frac{\{(V_{11}^{\eta} + V_{22}^{\eta})\eta_0 c + (V_{11}^{c} + V_{22}^{c})c - 2u_m\}(\tilde{B}_{11} + \tilde{B}_{22})}{2(s_{11} + s_{12})} + \frac{\{(V_{11}^{\eta} - V_{22}^{\eta})\eta_0 c + (V_{11}^{c} - V_{22}^{c})c\}(\tilde{B}_{11} - \tilde{B}_{22})}{2(s_{11} - s_{12})}\right] c^2|\tilde{\eta}|^2 + \left[\frac{(\tilde{B}_{12})^2}{2s_{44}} + \frac{(\tilde{B}_{11} - \tilde{B}_{22})^2}{4(s_{11} - s_{12})} + \frac{(\tilde{B}_{11} + \tilde{B}_{22})^2}{4(s_{11} + s_{12})}\right] c^4|\tilde{\eta}|^4 + \frac{\{(V_{11}^{\eta} - V_{22}^{\eta})\eta_0 c + (V_{11}^{c} - V_{22}^{c})c\}^2}{4(s_{11} - s_{12})} + \frac{\{(V_{11}^{\eta} + V_{22}^{\eta})\eta_0 c + (V_{11}^{c} + V_{22}^{c})c - 2u_m\}^2}{4(s_{11} + s_{12})} \quad \text{(A.19)}$$

Finally, for the case of $k_2 = 0$ and $\eta_0 = 0$ the evident form of the free energy renormalization under misfit stress appearance is

$$\Delta\tilde{F}_{FE} = \left(g - \frac{(2u_m - (V_{11}^{c} + V_{22}^{c})c)(B_{11} + B_{12})}{2(s_{11} + s_{12})} + \frac{(V_{11}^{c} - V_{22}^{c})(B_{11} - B_{12})}{2(s_{11} - s_{12})}c\right) c^2 k_1^2|\,\tilde{\eta}|^2 + \left(g - \frac{(2u_m - (V_{11}^{c} + V_{22}^{c})c)B_{12}}{s_{11} + s_{12}}\right)|\,\tilde{\eta}|^2 c^2 k_3^2 + \left(\frac{1}{4}\frac{(B_{11} + B_{12})^2 c^4}{s_{11} + s_{12}} + \frac{1}{4}\frac{(B_{11} - B_{12})^2 c^4}{s_{11} - s_{12}}\right) k_1^4|\,\tilde{\eta}|^4 + \left(\frac{(B_{12})^2 c^4}{s_{11} + s_{12}}\right) k_3^4|\,\tilde{\eta}|^4 + \left(\frac{(B_{11} + B_{12})B_{12}c^4}{s_{11} + s_{12}}\right) k_1^2 k_3^2|\,\tilde{\eta}|^4 + \frac{\{(V_{11}^{c} - V_{22}^{c})c\}^2}{4(s_{11} - s_{12})} + \frac{\{(V_{11}^{c} + V_{22}^{c})c - 2u_m\}^2}{4(s_{11} + s_{12})} \quad \text{(A.20)}$$

Note that the stabilization term proportional to fourth degree of order parameter is allowed by the symmetry and thus can be included in the free energy from the beginning. However, the material coefficient before the term is unknown and should still be defined. That is why we used the above way to re-derive the term with already defined coefficient from the entropy term. Actually, expansion on small η gives us that:

$$\frac{c\,k_B T}{2}\left[(1-\eta)\ln(1-\eta)+(1+\eta)\ln(1+\eta)\right]\approx\frac{c\,k_B T}{2}\left[\eta^2+\frac{\eta^4}{6}+O[\eta]^6\right] \quad (A.21)$$

The additional fourth power of η in Eq.(8a), $\gamma_{pr}=\frac{12\tilde{B}_{22}^2}{s_{11}+s_{12}}$, is related to the elimination of the elastic variables, namely elastic stresses and strains, from the initial free energy [given by e.g. above equations (A.20) and Eq.(5b)]. Mathematically, explicit dependences on the free energy coefficients and the order parameter η were obtained for stress and strains. These dependences were substituted into free energy (taking into account the Legendre transformation), which gave the renormalized energy (8b) and (9b).